\pdfoutput=1

\documentclass[aip,rsi,reprint]{revtex4-1}

\usepackage{lipsum}
\usepackage[usenames, dvipsnames]{color}
\definecolor{fade}{rgb}{0.9,0.9,0.9}

\usepackage{graphicx}
\usepackage{upgreek}
\usepackage{bm}

\usepackage{amssymb}
\usepackage{amsmath}

\renewcommand{\text}[1]{\ensuremath{\textrm{#1}}}
\newcommand*\diff{\mathop{}\!\mathrm{d}}

\begin{document}

\title{Spectral analysis and parameter estimation in levitated optomechanics}

\author{Chris Dawson}
\author{James Bateman}
\email{j.e.bateman@swansea.ac.uk}
\affiliation{Department of Physics, College of Science, Swansea University, SA2 8PP, UK}

\date{\today}

\begin{abstract}
  Optical levitation of nanoscale particles has emerged as a platform for precision measurement.
  Extremely low damping, together with optical interferometric position detection, makes possible exquisite force measurement and promises low-energy tests of fundamental physics.
  Essential to such measurement is an understanding of the confidence with which parameters can be inferred from spectra estimated from the indirect measurement provided by interferometry.
  We present an apparatus optimized for sensitivity along one motional degree of freedom, a theoretical model of the spectrum, and maximum likelihood estimation.
  The treatment accounts for the sinusoidal dependence of interferometric signal on particle position,
  and we use the technique to extract thermodynamic quantities in a regime where simpler treatments are confounded.
\end{abstract}

\maketitle


\section{Introduction}

Levitated optomechanics uses a nanoparticle, trapped in vacuum by the optical dipole force, as a harmonic oscillator in a thermal bath~\cite{bhattacharya2017levitated}.
Despite the apparent simplicity, interferometric position readout~\cite{gieseler2012subkelvin} combined with optical control, including intensity modulation~\cite{gieseler2012subkelvin,vovrosh2017parametric} and shaping of the focus~\cite{ricci2017optically},
yields a rich platform for exploring nanoscale dynamics~\cite{gieseler2018levitated,jain2016direct}, and promises ultra-precise measurements~\cite{ranjit2016zeptonewton,hebestreit2018sensing} and low energy tests of fundamental physics~\cite{geraci2010short,bateman2014nearfield,wan2016free,kaltenbaek2016macroscopic}.

Properties of the oscillator are not measured directly, but rather are inferred from optical measurements, often with the intermediate step of estimating spectral density from time-series data.
This optical measurement is interferometric and therefore not linear with position:
the linear approximation is appropriate for geometries where forward scattered light is used~\cite{gieseler2012subkelvin},
and care must be taken when extracting parameters in backwards scattered arrangements~\cite{mestres2015cooling,rashid2016experimental,vovrosh2017parametric}, where phase to position sensitivity can be much greater and there may be a non-zero phase offset.
Backscatter is desirable because of increased sensitivity and separation from the laser light, but even when used at low centre of mass temperatures, where the phase excursion is small, the offset can render measurement non-linear.

Furthermore, spectral overlap of oscillator modes, cross-coupling, and effects not accounted for in this simplistic description, notably rotation~\cite{monteiro2018optical,rashid2018precession,reimann2018ghz}, can pollute the spectrum and affect parameters extracted by, for example, integrating a truncated region of the spectrum.
For example, temperature, or a quantity proportional to it, can be found by integrating the spectral density associated with a given mechanical mode~\cite{clerk2010introduction},
but this is only possible when the mode is spectrally resolved,
precluding use of this simple technique at high pressure ($\gtrsim 10\,\textrm{mbar}$).

We describe an experimental system which is optimized for detection of one motional degree of freedom (longitudinal) while rejecting, to first order, signal from the other two (transverse).
We use backscatter, which offers a larger phase-shift for a given displacement than does forward-scatter, but which in our experiment introduces an uncontrolled and slowly drifting offset phase-shift;
and we use an optical fibre system which collects a significant fraction of the scatter and which guides this, without diffractive loss, to detection electronics.
This collection and guiding offers alignment stability and contrasts strongly with free-space detection where small photodiode area and the comparatively large free-space beams means that much of the collected light is unused.
Using scatter more efficiently is essential if we are to approach the standard quantum limit in these systems~\cite{chang2010cavity}.

In this manuscript, we
describe our experiment,
present a theoretical spectrum, including narrow-band limit and comparison with numerics,
and then apply a maximum likelihood approach to parameter extraction from simulated and experimental data.
Through this, we observe a centre of mass heating effect at intermediate pressure, where spectral overlap confounds simpler methods,
and finally we describe the experimental limits which we encounter in our specific implementation.

\section{Experimental apparatus}
\begin{figure}
  \centering
  \includegraphics[width=\linewidth]{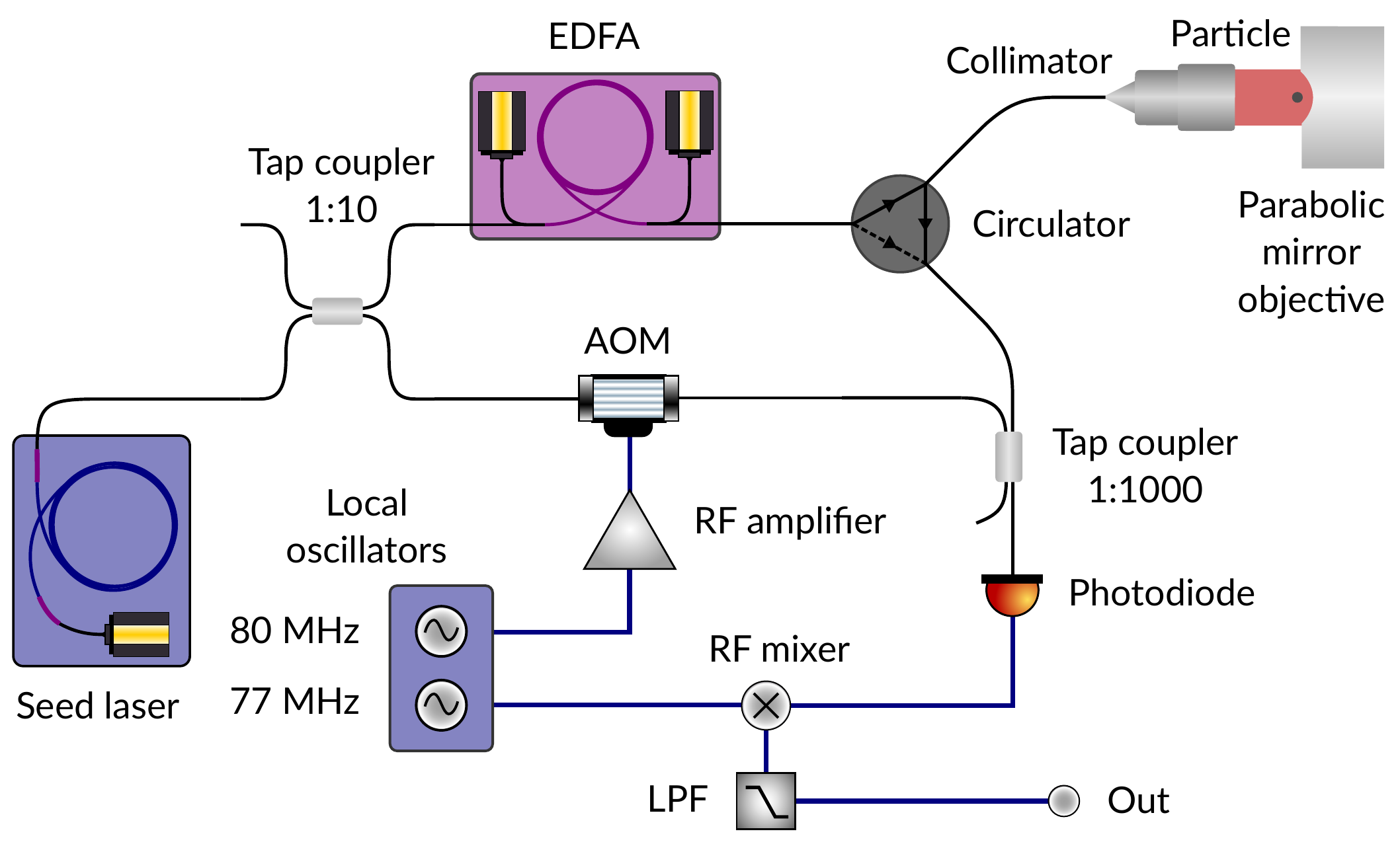}
  \caption{Experimental schematic showing light from the the low power seed laser split by a ratio 1:10, amplified by an erbium-doped fibre amplifier (EDFA), and directed towards the focussing mirror via a circulator.
  The optical fibres and components, aside from the seed laser, are non-polarization maintaining.
    The alternative path of this seed light goes via an acousto-optical modulator (AOM) which shifts the light in frequency by $80\,\textrm{MHz}$.
    The paths recombine with a ratio 1:1000 so that the majority of scattered light is retained.  The RF signal resulting from the photodiode detection is mixed with a reference near $80\,\textrm{MHz}$ and then low-pass filtered (LPF) before recording.}
  \label{fig:schematic}
\end{figure}

Our apparatus, illustrated in figure~\ref{fig:schematic}, consists of a single glass nanoparticle held, by the optical dipole force, in the focus of hemispherical parabolic mirror.
Gaussian beam illumination of this mirror is provided by a triplet collimator attached to a single-mode optical fibre.
Laser light is provided by a narrow linewidth telecommunications wavelength ($\lambda = 1550\,\textrm{nm}$) all-fibre laser, which is amplified by an erbium-doped fibre amplifier and sent, via an optical circulator, to the output collimator.
The trapped particle explores $\lesssim\lambda/10$ around the focus, and the potential is therefore well-approximated, for our purposes, as harmonic, with a different natural frequency along each of the Cartesian axes.

The particle Rayleigh scatters a small fraction of the trapping light, half of which is backscattered towards the parabolic mirror.
This light is collimated by the mirror and directed towards the fibre optic collimator, some fraction of which is then coupled into the optical fibre.
Unscattered light, and forward scattered Rayleigh light, diverges strongly as it travels towards the collimator, and only a small fraction of this is coupled back into the fibre.
Imperfections in components of the fibre optic network mean that other stray reflections, notably from the fibre output facet ($\sim 10^{-6}$) and the optical circulator ($\sim 10^{-5}$), make their way, with varying amplitudes and phases, to the photodiode.
Overall, light at this frequency, which in other schemes would provide a reference for interferometry, drifts in phase and amplitude because of the macroscopic path difference.
This motivates our introduction of phase-coherent light at a shifted frequency, implemented by the acousto-optical modulator (AOM) and known as heterodyne detection, as discussed in detail below.

This arrangement differs in two important ways from the forward-scatter, free-space detection scheme,
in which laser light and forward-scattered Rayleigh light falls onto a quadrant detector, with overall signal giving longitudinal information and the difference between left/right or top/bottom pairs giving transverse position information.
Firstly, in this backscatter scheme the optical phase of scattered light changes more rapidly with particle position along the optical axis;
secondly, the geometry and aggressive spatial filtering by the optical fibre mean that, when correctly aligned, this scheme is first-order insensitive to particle motion transverse to the optical axis.
For some applications in precision measurement, this larger sensitivity to one axis with strong rejection of others could be a strong advantage.
The fibre optic approach also means that, once coupled into the fibre, there are no diffractive losses and a large fraction of guided light can be directed onto a fast photodiode.

\subsection{Interferometric detection}
The photodiode signal $V$ arises from the interference of a reference field $E_\textrm{ref}$ (phase $\theta$) and scatter from the particle $E_\textrm{sca}$ (phase $\phi$).
The signal is proportional to the modulus squared of the total field, and contains offset (proportional to the sum of the squares of the individual fields) and a term sinusoidal with phase difference $\theta-\phi$ i.e.
$V\propto E_\textrm{ref}^2+E_\textrm{sca}^2+2E_\textrm{ref}E_\textrm{sca}\cos{\left(\theta-\phi\right)}$.

We model the phase shift from particle displacement as linear $\phi=\kappa z$ where the sensitivity $\kappa$ depends on the geometry, discussed below, and $z$ is particle position, along the optical axis, relative to the focus.
In the forward-scattered case, $\theta=-\pi/2$ via the Gouy shift, expansion to first-order in $\phi$ is justified because the overall phase excursion is typically small, and hence $V$ is approximately linear with position.
For backwards scatter, $\theta$ may drift, and phase excursion is not necessarily small.

\subsection{Position to phase sensitivity}
\label{sec:position-to-phase}
Through the focus of a Gaussian beam, the optical phase evolves more slowly than would a comparable non-focussed beam.
This Gouy shift, near the focus, modifies the rate of change of phase from $k=2\pi/\lambda$ to $k'=k-1/z_R$ where $z_R$ is the Rayleigh range.
Hence, for forward scatter, the rate of change (with particle position) of phase difference between Rayleigh scatter and laser field is $\kappa\approx k-k'=1/z_R$.
For back scatter, laser light accrues phase delay while travelling to the particle, and accrues additional phase delay on its return, giving the larger sensitivity $\kappa\approx k+k'=2k+1/z_R$.

When focussing with numerical aperture $\gtrsim 0.5$, the paraxial approximation underlying Gaussian beam treatments is not valid~\cite{varga2000focusing}.
For uniform illumination of a circular aperture~\cite{pang2011phase}, with numerical aperture $\textrm{NA}=1$ as in our parabolic mirror, we find $\kappa=2k-1.41\pi/\lambda$.
However, illumination would better be approximated by a truncated Gaussian, and we have not modelled the reflection and collimation of scattered light.

Experimental studies of similar parabolic mirrors have revealed strong sensitivity to alignment and manufacturing tolerances~\cite{bahk2005characterization,alber2017focusing}.
For lower aperture optical systems, or those employing compound lens microscope objectives, we could be more confident of the calculated value, and thereby convert the extracted phase modulation depth into a measurement of the temperature to mass ratio.
The proportionality of phase modulation to temperature remains, and we use this result later in this work.

\subsection{Heterodyne detection}
Interferometric detection in levitated optomechanics is usually by {\it homodyne} detection, where scatter interferes with light at the same frequency, i.e. $\theta$ is constant.
Here, implementation of this scheme is complicated by the presence of several contributions to light at this frequency, and a consequent drift in phase and amplitude because of the macroscopic (metre-scale) optical paths.

Instead, we employ {\it heterodyne} detection, where scattered light is interfered with phase-coherent, frequency-shifted light.
In our experiment, this light is derived from the same laser and frequency shifted using an acousto-optical modulator at $80\,\textrm{MHz}$.
It is guided to the photodiode and mixed with the scattered light using a 1:1000 coupler, which allows us to retain $\sim 99.9\%$ of the scattered light at this component.
The electrical signal from the photodiode is shifted down in frequency to $3\,\textrm{MHz}$ using a radio frequency components and $77\,\textrm{MHz}$ local oscillator.
Radio frequency mixing and filtering in any realistic device degrades the signal to noise;
implementations without this stage are possible using, for example, a electro-optical phase shifter at $3\,\textrm{MHz}$.

The principle advantage of heterodyne detection in our experiment is that spectrum remains stable regardless of slow drifts in the offset of the phase $\theta$, which now evolves at a rate fast compared with the dynamics of the system.
It also allows us to use results from radio-frequency communication to understand the spectrum.

\section{Theoretical spectrum}
For illustration, although we will soon improve this description, we consider purely harmonic particle motion $z=z_0\sin(\Omega t)$, which gives phase modulation $\phi=\phi_0\sin(\Omega t)$ where $\phi_0 = \kappa z_0$, with a phase offset $\theta$ treated as constant over the timescale of observation.  We find
\begin{eqnarray}
  \label{eq:jacobianger}
  &&\cos\left[\theta-\phi_0\sin\left(\Omega t\right)\right]=\cos\theta J_0 + \nonumber\\
  &&2\cos\theta\sum_{n\geq2}^\textrm{even}J_n\cos(n\Omega t) + 2\sin\theta\sum_{n\geq1}^\textrm{odd}J_n\sin(n\Omega t)
\end{eqnarray}
where $J_n$ is the n$^\textrm{th}$ order Bessel function evaluated at $\phi_0$.
This description has been used to understand optomechanical spectra, including a case where $\theta$ was varied systematically~\cite{vovrosh2017parametric}.
The (co)sinusoidal dependence of (even)odd harmonics is a feature of this expansion being around zero frequency, and a useful interpretation is of negative orders being reflected about the origin to overlap and interfere with their positive frequency counterparts.

We now improve the description of particle motion.
To obtain a spectrum, we necessarily observe for a time long compared with the relaxation time.
Therefore, it is not accurate to treat motion as purely harmonic.
Hereafter, we treat the particle as a stochastic harmonic oscillator, which leads to an important difference. 
Moreover, to avoid interference of negative and positive orders, we expand about a frequency large compared with the width of the spectrum, by using $\theta=\omega_0 t+\theta_0$.

The differential equation describing a stochastic harmonic oscillator, with natural frequency $\Omega$ and damping $\Gamma$, is
\begin{equation}
  \ddot{z}+\Gamma\dot{z}+\Omega^2z=w(t)
\end{equation}
where $z$ is the particle position, over-dot means differential with time, and $w$ is a Wiener process with volatility $k_BT\Gamma/M$ chosen to agree with equipartition; $k_B$ is Boltzmann's constant, $T$ is centre of mass temperature, and $M$ is the particle mass.
From the fluctuation--dissipation theorem and linear response theory, the spectrum of fluctuations is
\begin{equation}
  \label{eq:Szz}
  S_{zz}(\omega)=\frac{2k_BT}{M}\,\frac{\Gamma}{\left(\omega^2-\Omega^2\right)^2+\Gamma^2\omega^2}.
\end{equation}

For forward scatter at low temperature, where $\theta=-\pi/2$ and $\phi_0\ll 1$, detection is approximately linear and it is often sufficient to approximate the signal as a scaled version of equation~\ref{eq:Szz}.
For backward scatter, where these conditions are not satisfied, previous
work has combined equations~\ref{eq:jacobianger} and \ref{eq:Szz} in a heuristic way, by considering a scaled delta function at each frequency, broadened by the position spectrum~\cite{vovrosh2017parametric}.
Here, instead, we use a result from radio communications which gives an exact result.

The \emph{correlation function} $R_{vv}(t)=\langle v(t)v(t-\tau)\rangle_\tau$ of a signal $v(t)=v_0\sin\left[\omega_0 t +\phi(t)\right]$ phase-modulated by a Gaussian random process $\phi(t)$ with correlation function $R_{\phi\phi}(t)$ and variance $\Phi^2=\kappa^2\langle z^2\rangle$ is~\cite{godone2008rf}
\begin{equation}
\label{eq:middletontime}
R_{vv}(t)=v_0^2\exp{\left[R_{\phi\phi}(t)-\Phi^2\right]}
\end{equation}
the Fourier transform of which is the spectrum and can be represented as~\cite{middleton1951distribution,brochard2017power}
\begin{equation}
  \label{eq:middleton}
  \sigma_{vv}(f)=e^{-\Phi^2}\sum_{n=0}^\infty \frac{\Phi^{2n}}{n!}\sigma_{zz}(f)\overset{n}{\circledast}\sigma_{zz}(f)
\end{equation}
where
$\sigma_{zz}(f)=2\pi\,S_{zz}(2\pi f)/\langle z^2\rangle$ is the normalized position spectrum (such that $\int_{-\infty}^{+\infty}\sigma_{zz}(f)df=1$),
and $\overset{n}{\circledast}$ is the $n^\textrm{th}$ order convolution infix operator defined recursively via $a\overset{n}{\circledast}a = a\circledast a\overset{n-1}{\circledast}a$, with $a\overset{2}{\circledast}a=a\circledast a$ the standard convolution, $a\overset{1}{\circledast}a=a$, and $a\overset{0}{\circledast}a=\delta$ the Dirac delta.

The overall appearance is of peaks near integer multiples of natural frequency $\Omega$, as in equation~\ref{eq:jacobianger}.
Higher order terms give contributions to lower order harmonics, and hence the energy (and information) contained within a given harmonic is a result of a series summation,
not just the contribution from a given order.
This expression valid for under- and over-damped oscillators.

\subsection{Illustrative, typical spectra}
\begin{figure}
  \centering
  \includegraphics[width=\linewidth]{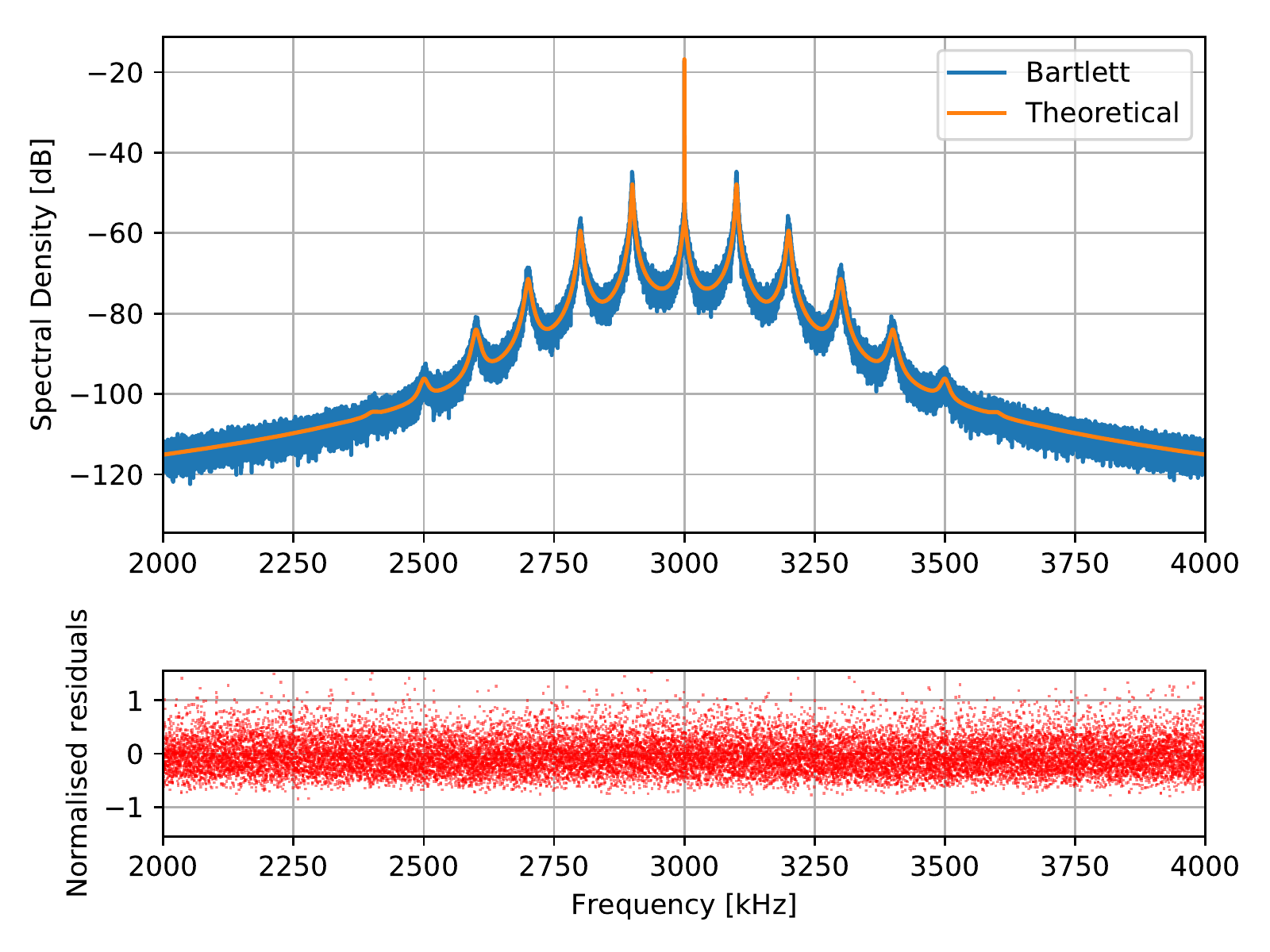}
  \caption{(a) Illustration of the theoretical spectrum (orange) compared with that obtained by simulating particle trajectories with stochastic differential equations and estimating the spectral density using Bartlett's method (blue) with 9 periodograms, with model parameters $\Omega=2\pi\,\times 100\,\textrm{kHz}$, $\Gamma=20\,\textrm{kHz}$, $\Phi=0.75$. (b) Normalised residuals of the same.}
  \label{fig:cf}
\end{figure}

An example of a typical spectrum is shown in figure~\ref{fig:cf}.
Also shown here is a spectrum estimated from a simulated realisation of a random process; spectral estimation is discussed further in Section~\ref{sec:estimation-from-time-series}.

The spectrum is centred on the chosen modulation frequency, $\omega_0=2\pi\,\times\,3\,\textrm{MHz}$, with peaks spaced at integer multiples of $\Omega$ symmetrically either side.
The amplitude of higher order peaks reduces monotonically and series convergence is assured because of the factorial in the denominator.

The overall, broad pedestal arises from the increasingly broad contribution of higher order convolutions to the overall spectrum.
The presence of this pedestal makes clear that the integrated area under the first order peak is not necessarily an accurate representation of the variance in $\phi(t)$,
especially because the {\it shape} of the spectrum will change as this variance is e.g. reduced by cooling.

\subsection{Narrow-band limit}
\begin{figure}
  \centering
  \includegraphics[width=\linewidth]{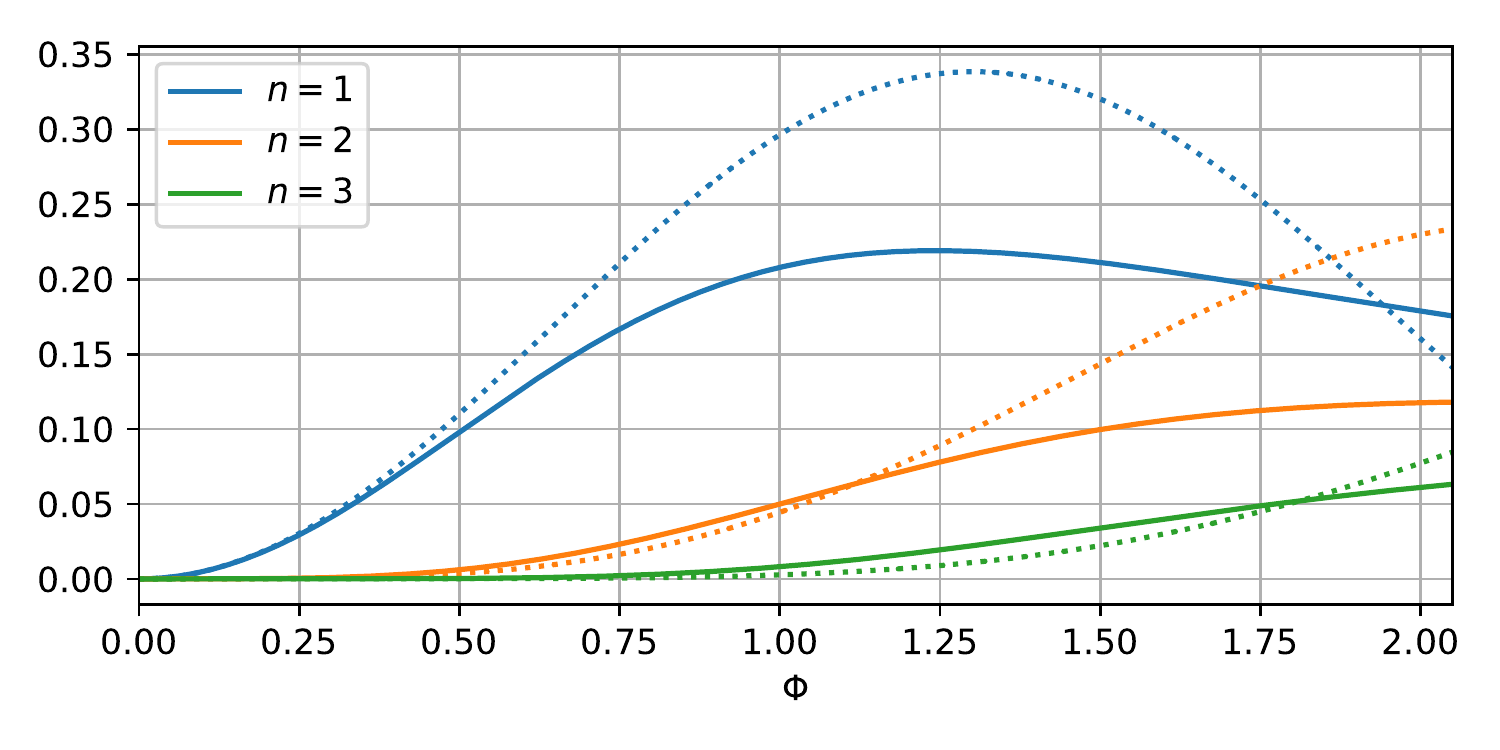}
  \caption{Relative fraction of signal contained within $n^\textrm{th}$ order peak, for the narrow-band case, according to Equation~\ref{eqn:correlation_narrow}, or using (unmodified) Bessel functions as might be expected when treating motion as purely harmonic.  The deviations are significant for large modulation depth. The variance of $\phi=\phi_0\cos\Omega t$ is $\phi_0^2/2$, and hence we plot $J_n(\sqrt{2}\,\Phi)^2$ (dashed) and $I_n(\Phi^2)$ (solid).}
  \label{fig:bessel-cf}
\end{figure}

Often, with this kind of system, one has a narrow-band process.  Modelling this as purely sinusoidal gives Equation \ref{eq:jacobianger}; instead, we
observe that the limiting case for low damping is that the process has 
a sinusoidal \emph{correlation function} $R_{\phi\phi}=\Phi^2\cos\Omega t$ for which, using Equation~\ref{eq:middletontime} and the Jacobi--Anger identity with imaginary amplitude, we find
\begin{equation}
\label{eqn:correlation_narrow}
R_{vv}^\textrm{(narrow)}(t)= v_0^2 e^{-\Phi^2}\left[I_0 + 2\sum_{n=1}^\infty I_n\cos{n\Omega t}\right]
\end{equation}
where $I_n$ is the modified Bessel function of the first kind evaluated at $\Phi^2$.
Details of the algebra are given in Appendix~\ref{sec:narrow-band}.
Previous work\cite{mestres2015cooling,vovrosh2017parametric} used the (non-modified) Bessel functions to estimate physical parameters from observed spectra, which agrees with Eq.~\ref{eqn:correlation_narrow} for small $\Phi$.

The ratio of predicted peak amplitudes, relative to the first order, is illustrated in figure~\ref{fig:bessel-cf}, and the treatments agree for small $\Phi$.
Notably, treating motion as purely harmonic predicts that the first order will vanish at $\Phi\approx 3.8$, the first zero of $J_1$; the modified Bessels have no such zero crossings.

\section{Parameter estimation from time series}
\label{sec:estimation-from-time-series}
Our goal is to estimate model parameters describing the process from the time-series measurements of a phase-modulated signal arising from a realisation of this process.
Ideally, we would compute the likelihood of the time-series data for given model parameters, and thereby infer the probability density for these parameters~\cite{sivia2006data}.
However, the experimental spectrum, while relatively clean, contains features not described by this model, and these are easily filtered in the spectral domain.
Therefore, we first make a non-parametric estimate of the spectrum, and then compute the likelihood of this spectrum, over the relevant regions, for given model parameters.
Future work may address this parametric estimation problem without the intermediate non-parametric spectral estimation step.
There is interest in this approach, but to our knowledge the existing treatments (e.g. autoregressive maximum likelihood; other Bayesian methods~\cite{singh2018fast}) are not applicable when the measurement of the process is non-linear.

We base our spectral-domain approach on Whittle's approximate log-likelihood~\cite{taniguchi2000asymptotic}: 
\begin{equation}
\mathcal{L}(\bm{\alpha})=\sum_i \log S(\bm{\alpha}) + \hat{S}/S(\bm{\alpha})
\end{equation}
where $\bm{\alpha}$ are the model parameters, $S$ is the theoretical spectrum, and $\hat{S}$ is an estimator of the spectrum from the time-series. 
Summation is over the discrete frequencies at which the spectrum is estimated.
The probability density for the parameters given the data~\cite{sivia2006data} is $\textrm{prob}(\bm{\alpha}|\hat{S})\propto \exp\left[-\mathcal{L}(\bm{\alpha})\right]$.

The Whittle likelihood is an approximation of the true likelihood for a stationary Gaussian time-series model.
While the underlying harmonic oscillator can be described by such a model, our measurement of it cannot.
Therefore, it is unclear whether the Whittle likelihood will give an accurate estimate in this case.
We make a slight adaption (described below) and verify the effectiveness numerically in our use case.

\begin{figure}
  \centering
  \includegraphics[width=\linewidth]{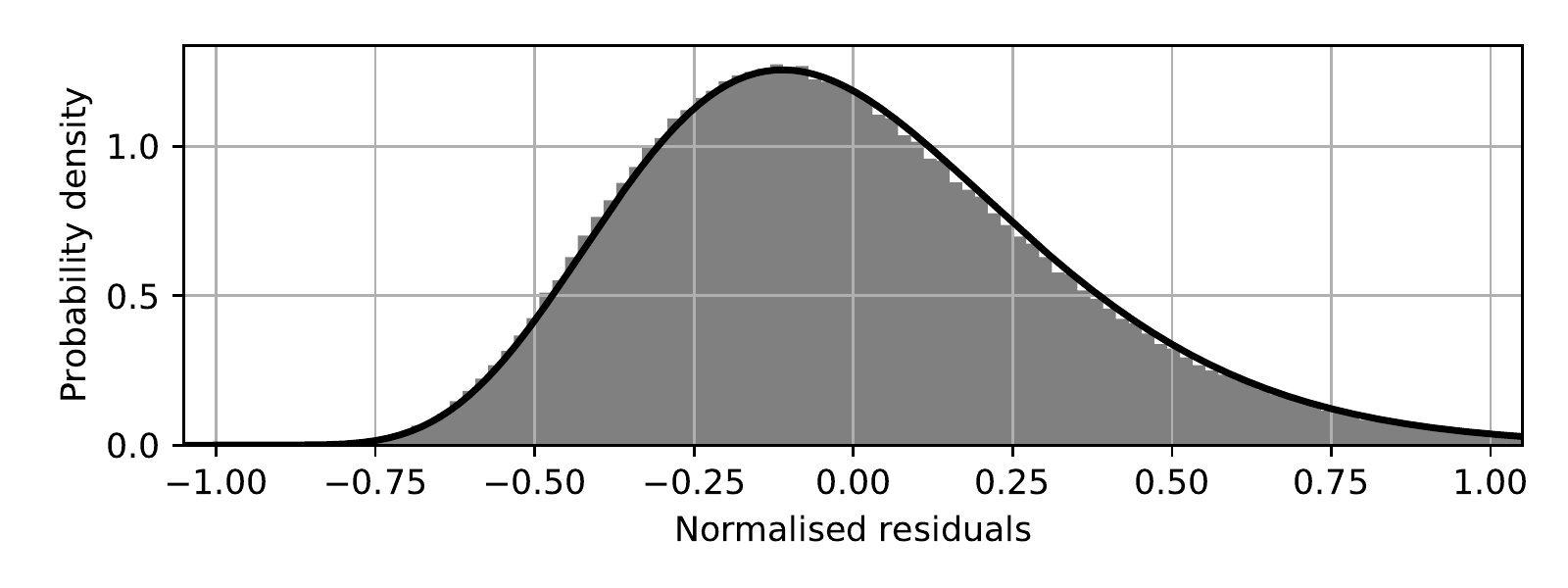}
  \caption{A histogram of normalised residuals from figure~\ref{fig:cf}(b) (grey), with a theoretical $\chi^2_\nu$ distribution (black).  For these numerics, to illustrate the $\chi^2$ nature and the tend towards Gaussian, we use the relatively low value of $\nu=18$.}
  \label{fig:cf-hist}
\end{figure}

The Whittle likelihood was originally formulated for the periodogram, the modulus squared of the discrete Fourier transform.
This estimator is asymptotically unbiased, but it is not consistent: the variance does not decrease for a large number of points.
We trade points for consistency by averaging the periodogram using Bartlett's method:
split the time-series data of length $N$ into many $M$-length segments, each of which is windowed using the Tukey--Hanning window, and then compute the average periodogram over these segments.
The distribution of this estimator relative to the true value is~\cite{priestley1981spectral} 
$\nu\hat{S}/S\sim \chi^2_\nu$ where this is the $\chi$-squared distribution, and the degrees of freedom $\nu$ is twice the number of segments, $\nu=2N/M$;
this is illustrated in figure~\ref{fig:cf-hist}.
For large $\nu$, this tends to a Gaussian distribution $\hat{S}/S - 1 \sim \mathcal{N}(0,\sqrt{2/\nu})$

\subsection{Parameter estimation with simulated trajectories}
\label{sec:estimation-with-simulated}
\begin{figure}
  \centering
  \includegraphics[width=\linewidth]{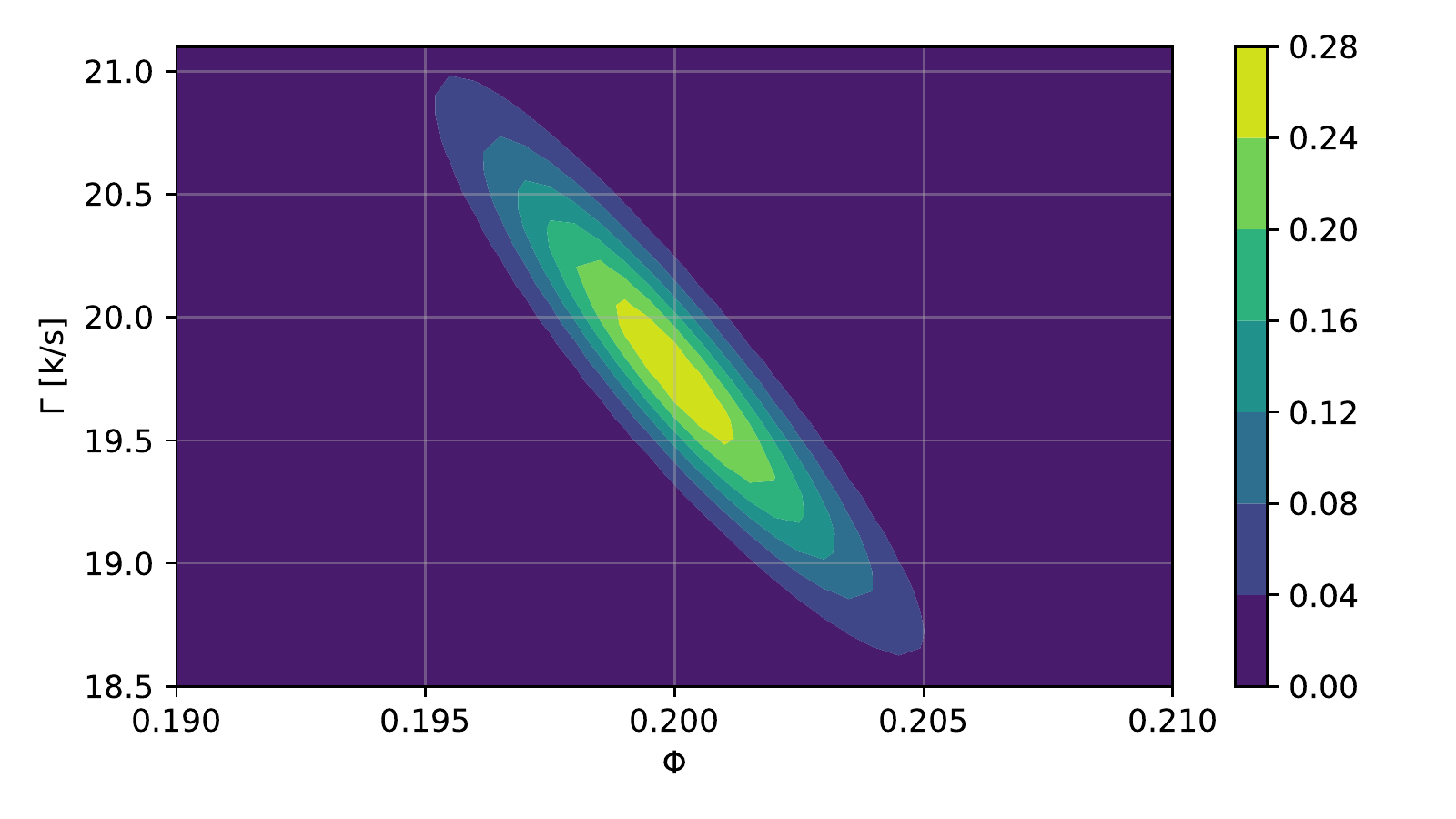}
  \caption{Probability density obtained from $\mathcal{L}$ by varying $(\Phi,\Gamma)$ for $\Omega=2\pi\,\times\,100\,\textrm{kHz}$.  The natural frequency is well constrained so it is reasonable to compute at fixed $\Omega$ rather than marginalising.  The true values $(\Phi=0.20, \Gamma=20\,\textrm{kHz})$ are contained within the uncertainty ellipse, and the anti-correlation between these parameters is apparent.}
  \label{fig:whittle}
\end{figure}

We compute the likelihood function and consequent probability density for a simulated trajectory with known parameters $\bm{\alpha}_0$, as we vary $\bm{\alpha}$ about the true value.
The physical parameters describing our spectrum are $\bm{\alpha}=(\Phi,\Omega,\Gamma)$.
For illustration, since $\Omega$ is well constrained, we vary $(\Phi,\Gamma)$ while holding $\Omega=\Omega_0$.
Details of the simulation are described in Appendix~\ref{sec:ensemble-simulation}, and the calculated probability density is illustrated in figure~\ref{fig:whittle}.
The central spectral peak, the delta function resulting from $n=0$ in equation~\ref{eq:middleton}, contains no information about the random process, is relatively large, and depends critically on the spectral windowing function;
we therefore exclude a small region near to this peak from our likelihood calculations.

To assess whether this Whittle probability density is an accurate representation of the information which can be extracted through this process, we
we create an ensemble of simulations with known $\bm{\alpha}$, find the maximum likelihood estimate in each case $\bm{\alpha}_\textrm{MLE}$, and estimate properties of the assumed Gaussian probability distribution from which these estimates are picked.
This ensemble estimate of the probability distribution is compared with that obtained by computing $\exp\left[-\mathcal{L}(\bm{\alpha})\right]$ directly.

Rather than raster the parameter space, we compute the \emph{profile likelihood}, by constraining one element of $\bm{\alpha}$ and fitting all others.
This gives access to the marginal probability, effectively integrating over the other (nuisance) parameters.
The difference between computing marginal probability density $\textrm{prob}(\Phi)=\int \textrm{prob}(\Phi,\Gamma)\, \diff\Gamma$ and evaluating $\textrm{prob}(\Phi,\Gamma=\Gamma_0)$ is apparent in figure~\ref{fig:whittle}, where the extremal $\Phi$ accessible for a given probability density is larger if we integrate over $\Gamma$ (marginal distribution) or allow $\Gamma$ to be adjusted (profile likelihood),
compared with constraining $\Gamma=\Gamma_\textrm{MLE}$.

\section{Application to experimental data}

Experimental measurements are subject to additional complications not captured by the simulations.
The collected optical power and the responsivity of the photodetector give some scaling to the recorded voltage signal, and an advantage of the technique is that information is encoded in the spectral {\emph shape}, not the absolute scale.
In addition, the detection system introduces measurement noise, which we treat as white, and which therefore manifests spectrally as a constant offset.
We treat this scaling and offset as nuisance parameters by finding their maximum likelihood values for each calculation of $S(\bm{\alpha})$, thereby computing the profile likelihood as described in Section~\ref{sec:estimation-with-simulated}.

\begin{figure}
  \centering
  \includegraphics[width=\linewidth]{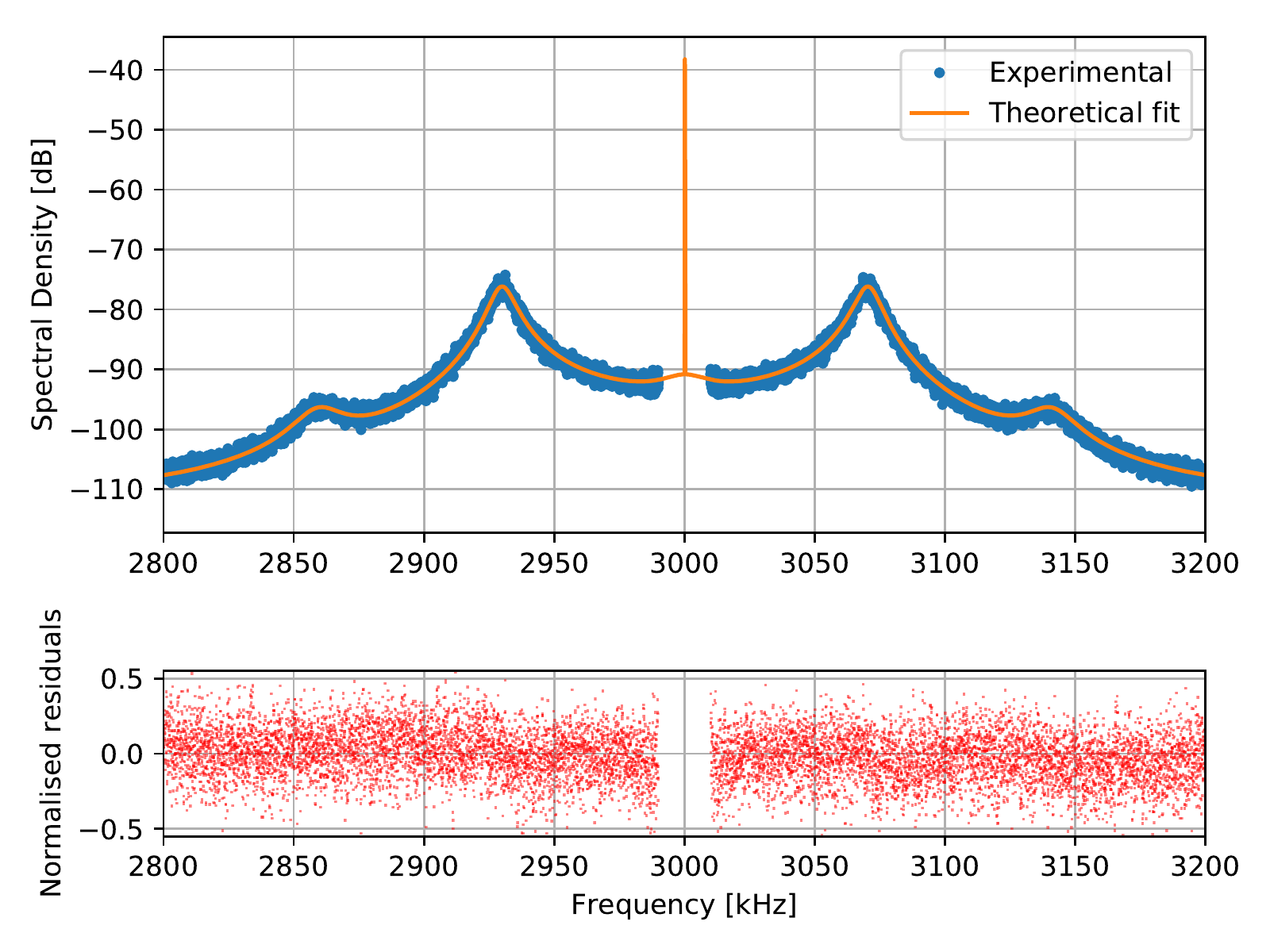}
  \caption{(a) Illustration of the fitted theoretical spectrum (orange) compared with that obtained experimentally by recording time-series and estimating the spectral density using Bartlett's method (blue) with 38 periodograms, with best-fit parameters $\Omega=2\pi \times 70.5\,\textrm{kHz}$, $\Gamma=62.0\,\textrm{kHz}$, $\Phi=0.24$. (b) Normalised residuals of the same. The central peak is excluded because this contains no information about the motion, the amplitude drifts from multi-path interference, and the width is dominated by the spectral windowing function.}
  \label{fig:cf-exp}
\end{figure}

An example of a typical maximum likelihood fit at a reasonably high pressure, where overlap between peaks is significant, is shown in figure~\ref{fig:cf-exp}.
Normalised residuals are shown and, because we are averaging over a large number of periodograms, their distribution approaches Gaussian.

\subsection{Probability densities and parameter uncertainties}

\begin{figure}
  \centering
  \includegraphics[width=\linewidth]{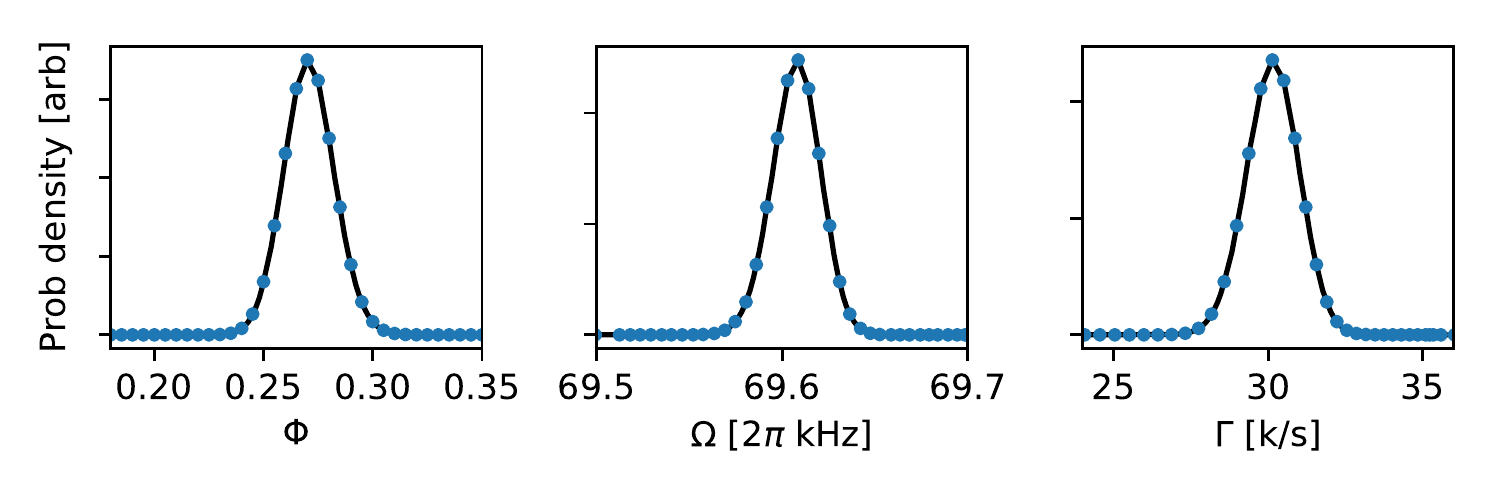}
  \caption{Example of probability densities computed by the profile likelihood method of constraining (in this case) $\Phi$ and fitting all other parameters in $\bm{\alpha}$.  For scenarios where $\Omega$ and $\Gamma$ are well-behaved, the trajectory through parameter space reveals probability densities for both constrained and fitted parameters.}
  \label{fig:prob-density}
\end{figure}

Once parameters $\bm{\alpha}_\textrm{MLE}$ have been found which maximize the Whittle likelihood, we
compute the probability density for a given parameter by constraining this parameter (to some value near to the maximum likelihood value) and then maximizing probability by adjusting all other parameters.
Moreover, when the fitted parameters depend monotonically on the constrained parameter, then this calculation also reveals the probability density for the fitted parameters.
An example of probability densities computed in this way is shown in figure~\ref{fig:prob-density}.

For spectra with relatively high damping ($\Gamma\gtrsim 1\,\textrm{kHz}$) our model describes the experimental spectrum well, and we are able to extract $\Phi$ and the associated uncertainty in a regime where spectral overlap with harmonics would confound the naive approach of integrating under peaks.
For spectra with lower damping, the intensity stability of our experiment affects the peak shape, principally through the square-root dependence of $\Omega$ on power; this limitation is particular to our apparatus, does not limit the technique in general, and is discussed further in Section~\ref{sec:RIN}.
For $\Gamma\lesssim 150\,\textrm{Hz}$ we expect an effect from windowing of our finite-length time series, and this could be included in a description of the theoretical spectrum.

\subsection{Observing heating at intermediate pressure}
\begin{figure}
  \centering
  \includegraphics[width=\linewidth]{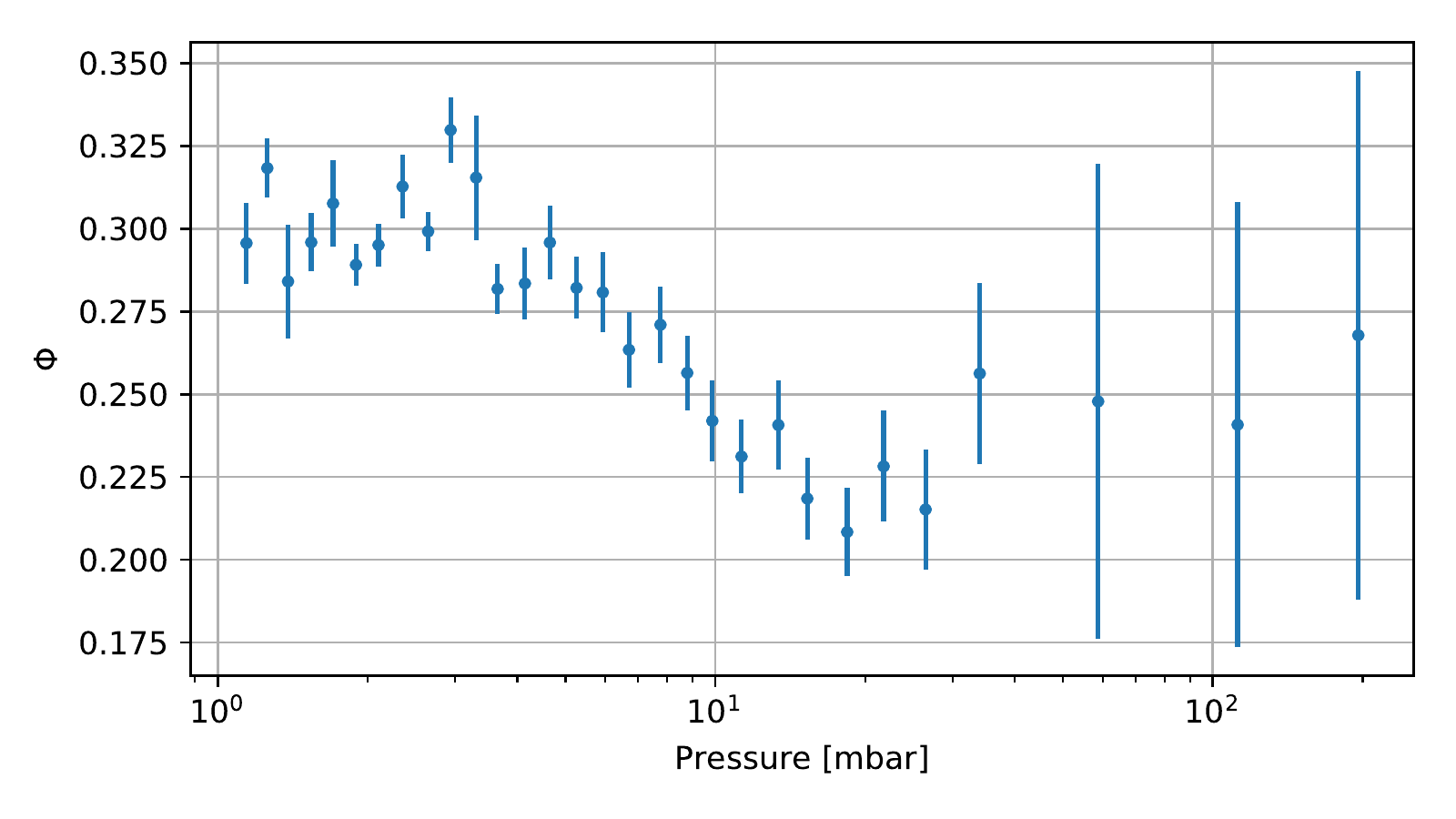}
  \caption{Experimentally observed phase modulation $\Phi$ with uncertainties derived from the Whittle likelihood.  $\Phi$ is not well determined above $100\,\textrm{mbar}$, with a probability density which is non-Gaussian and consistent with zero; at low pressure on this scale, the effect of laser intensity noise affects the estimate. Near $10\,\textrm{mbar}$, where this technique works well, we observe an increase in the phase modulation depth, which we associate with an increase in centre of mass temperature.}
  \label{fig:Phi-with-pressure}
\end{figure}
We apply our estimation technique to spectra obtained for a nominally $100\,\textrm{nm}$ silica particle in a dipole trap, as we reduce the gas pressure in the vacuum chamber.
Probability density for $\Phi$ is found by computing the Whittle likelihood as this parameter is constrained and all others are adjusted to maximize this likelihood.
The best-estimate and uncertainty are calculated from this probability distribution in the standard way, and results are plotted as a function of pressure in figure~\ref{fig:Phi-with-pressure}.
The three points at highest pressure (around $100\,\textrm{mbar}$) are derived from probability distributions which are not Gaussian, but are sufficiently broad so as to not be misleading when described by symmetric errorbars.

As discussed in Section~\ref{sec:position-to-phase}, the phase to position sensitivity is not known accurately in this system.
The phase modulation depth is proportional to temperature $\Phi^2\propto T/(M\Omega^2)$
and hence, because $\Omega$ is well-constrained and we assume particle mass $M$ remains unchanged,
we interpret the slope near $10\,\textrm{mbar}$ as an increase in temperature.
Comparing the relative plateaus near $20\,\textrm{mbar}$ and $2\,\textrm{mbar}$, we estimate an increase in temperature of approximately $90\,\textrm{K}$.

Decrease in mass would also manifest as an increase in $\Phi$.
Recent experiments on similar systems have revealed that micron-sized particles are porus and can contain significant water~\cite{blakemore2019precision}.
However, the strong absorption of $1550\,\textrm{nm}$ light by water suggests that there would be low content even at atmospheric pressure.
We cannot rule out the possibility that mass reduces, and future experiments, perhaps cycling pressure with different background gases, may be informative.

Some experiments in levitated optomechanics have suffered from an unexpected increase in particle loss probability at these intermediate pressures,
and there has been work to understand temperature in this setting\cite{millen2014nanoscale,hebestreit2018calibration}.
Experiments using telecommunications wavelength $1550\,\textrm{nm}$ (rather than  $1064\,\textrm{nm}$) appear to suffer less from this unexplained loss,
and this is assumed to be because of the lower material absorption of silica at this wavelength.
This technique may be a useful tool to estimate temperature changes in this regime, 
with applications including diagnosing material properties of fabricated nanoparticles designed to minimize heating caused by laser absorption\cite{frangeskou2018pure}.

\subsection{Intensity noise and non-linear broadening}
\label{sec:RIN}

An additional complication, relevant at low $\Gamma$, is relative intensity noise: the laser intensity at the focus has some small, low-frequency drift, and this affects properties of the spectrum.
For a fibre laser, the noise spectrum is extremely quiet at or above the particle oscillation frequency $\Omega$, making this a good choice to minimize parametric heating\cite{savard1997laser}; it is more significant at low frequencies, corresponding to slow drifts during data collection.

From direct measurement of the intensity at points in our fibre network we constrain $R\lesssim 1\%$ and therefore this is significant only for $\Gamma\lesssim \Omega/100\sim 10\,\textrm{kHz}$.

To model this effect, we assume that drifts in intensity are slow compared with the relaxation time of the oscillator $\Gamma^{-1}$ and that the distribution of intensity is Gaussian with some width $R$.
This model is unlikely to be sufficient for precision measurements, and future experimental work must be undertaken to minimize this drift.
Under these assumptions, we can describe the observed spectrum as an average over the intensity distribution.

The intensity affects several aspects of spectrum: the overall scale (linearly); the natural frequency $\Omega$ (square root); and the modulation depth $\Phi$ (square root) via the natural frequency because the spatial extent of the thermal state depends inversely on $\Omega$.  Therefore, the overall broadened spectrum is
\begin{align}\label{eq:RIN}
S'(\Phi,\Omega,\Gamma) = \frac{1}{\sqrt{2\pi}R} \int (1+r)\, e^{-r^2/(2R^2)}\times\nonumber\\
S(\Phi/\sqrt{1+r},\Omega\,\sqrt{1+r},\Gamma)\,\diff r\,.
\end{align}

\begin{figure}
  \centering
  \includegraphics[width=\linewidth]{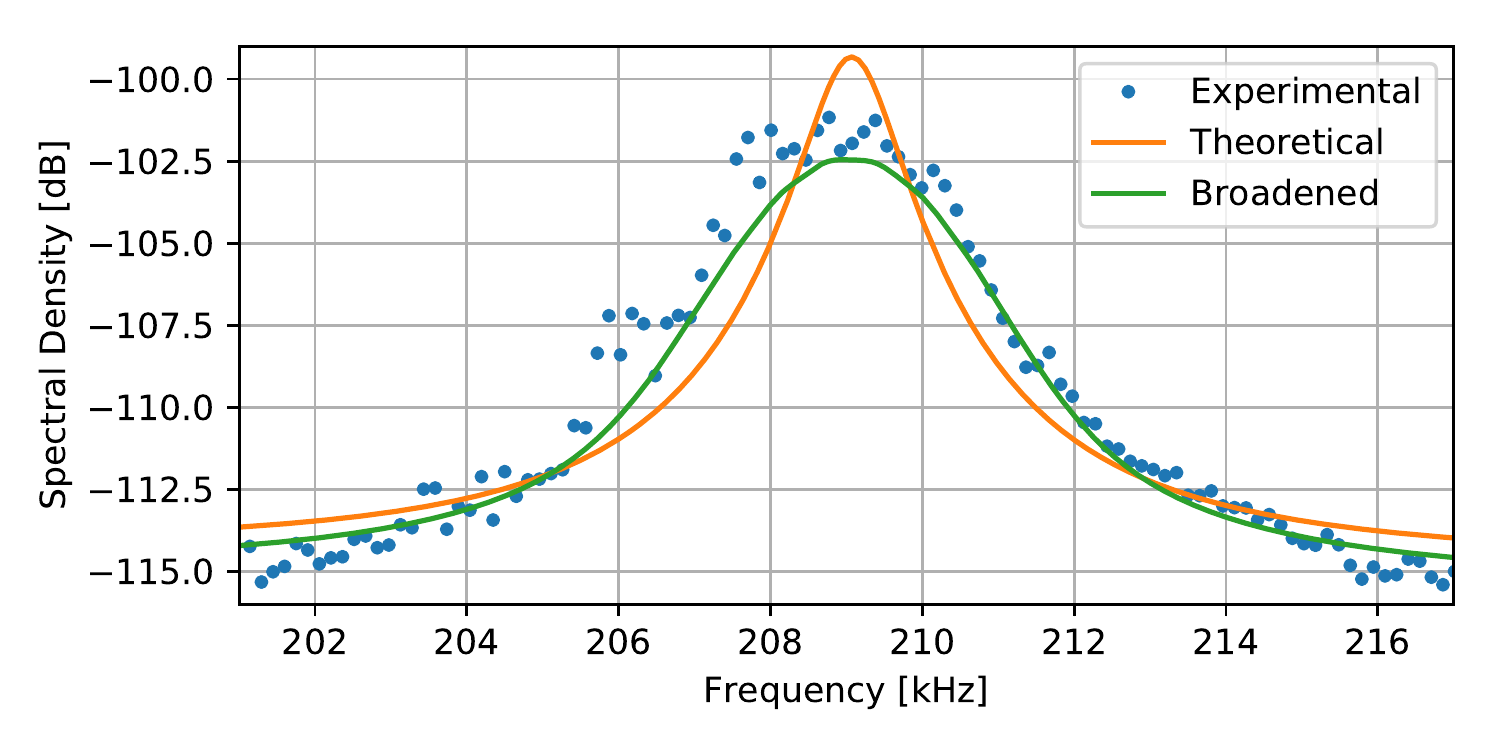}
  \caption{Comparison of experimental spectra with theoretical spectrum and a theoretical spectrum broadened by inclusion of relative intensity noise ($R=1\%$) illustrating that this is significant even for small intensity noise.  Other parameters are $\Phi=0.35$, $\Omega = 2\pi \times 69.8\,\textrm{kHz}$, and $\Gamma=2.5\,\textrm{k}/\textrm{s}$ and the pressure is $0.8\,\textrm{mbar}$, just below the minimum of figure~\ref{fig:Phi-with-pressure}.  The maximum likelihood fit is over the entire spectrum, and a zoomed region around the 3rd order peak is shown, where the broadening and damping are comparable i.e. $R\times 3\Omega\approx\Gamma$.}
  \label{fig:RIN-zoom-3rd}
\end{figure}

An example spectral peak, with experimental, unbroadened theoretical, and broadened theoretical according to equation~\ref{eq:RIN} is shown in figure~\ref{fig:RIN-zoom-3rd}.
The model parameters are found by maximizing the likelihood over the entire spectrum, and a zoomed in region, where the effect is most visible, is shown for illustration.
While this Gaussian broadening captures the behaviour well, and quite often fits our experimental results, care must be taken if this is used for parameter extraction
because the duration of the measurement is not sufficiently long that the intensity distribution can be reliably approximated as Gaussian; the dynamics of the drift are too slow.
Sometimes, for example, intensity undergoes a linear drift, which results in an {\it asymmetric} peak; this might explain examples in the literature, e.g. figure~3 in Ref.~\onlinecite{vovrosh2017parametric}.
Estimates based on the integrated area are unaffected, but the use of this whole spectrum method must account more carefully for any such drift.

An additional source of broadening is the thermal average of the Duffing non-linear frequency shift \cite{gieseler2013thermal}.  This effect, and the consequent distinctive asymmetric peak shapes, is masked in our system by the slightly larger intensity noise.  It is straightforward to include this non-linear broadening by averaging the heterodyne spectrum over the Boltzmann distribution, with frequency shift $\Delta\Omega\propto E$ and oscillation amplitude variance $\Phi\propto\sqrt{E}$, similarly to equation~\ref{eq:RIN}.  In constrast with slow intensity drifts, it is reasonable to sample sufficiently long that the Boltzmann average is a good approximation, and therefore we expect, although cannot currently verify, that our parameter estimation approach remains valid.

\section{Conclusions}
We have presented a technique for extracting, with confidence intervals, thermodynamic quantities from interferometric position measurements of a levitated nanoparticle by careful treatment of the estimated spectral density.
The techniques relies on the shape of the spectrum, and is indifferent to calibration of the photodiode responsivity or changes in the signal amplitude.
We have demonstrated this technique with experimental apparatus which is long-term stable and optimized for sensitivity along one direction, with strong rejection of others, giving a spectrum well described by the model.
The technique allows extraction when spectral features are not well resolved, and permits spectral windowing, which means it can be used when experimental spectra are cluttered by unmodeled features.

We have used this technique to observe centre of mass heating at intermediate pressure, where the simpler technique of integrating under a peak is not appropriate.
This technique may find use in diagnosing temperature dependence in this pressure range, for example when characterising unwanted heating in carefully fabricated extremely pure nanodiamonds.
Alternatively, it may be used if the temperature is known and the mass changing, such as by deliberate evaporation of a nanoparticle to obtain small trapped nanoparticles, with thermalisation via buffer gas.

This implementation was limited by intensity noise, but this is not a fundamental limitation of the technique.
Further, one could in principle record for sufficiently long that the histogram of intensity fluctuations is well described by a Gaussian, for which the model would then be expected to fit, but a reduction in the noise is a more efficient approach.

The high aperture optical trap, with possible manufacturing and experimental imperfections and an incomplete model of the focussing and collection optics, means that we cannot, with confidence, calculate the phase to position sensitivity in our experiment; if this were better known, by either direct measurement, calculation, or using different optics, the technique would allow for direct calculation of the temperature to mass ratio from the extracted phase modulation depth.

This work focussed on heterodyne detection, which can be implemented fully optically with little loss in signal quality.
However, many existing experiments use homodyne detection, and future work is to extend the formalism to cover this case.

\section*{Acknowledgements}
This work was supported by startup funding from the College of Science at Swansea University.
The authors would like to thank George Winstone for helpful discussions.

\bibliography{main}

\appendix
\section{Narrow band spectrum}
\label{sec:narrow-band}
Starting with equation~\ref{eq:middletontime} for the correlation function $R_{vv}$ of the phase-modulated signal $v$,
\begin{equation}
R_{vv}(t)=v_0^2\exp{\left[R_{\phi\phi}(t)-\Phi^2\right]}
\end{equation}
we use a narrow-band process $\phi$ which has correlation function $R_{\phi\phi}=\Phi^2\cos{\Omega t}$ to find
\begin{equation}
R_{vv}(t)=v_0^2e^{-\Phi^2}e^{-\Phi^2\cos(\Omega t)}.
\end{equation}

The modulated exponential term can be expressed using the Jacobi--Anger identity with the replacements $iz=\Phi^2$ and $\theta=\Omega t$:
\begin{equation}
  e^{iz\cos\theta}=J_0(z)+2\sum_{n=1}^\infty i^nJ_n(z)\cos(n\theta)
\end{equation}
where $J_n$ is the $n^\textrm{th}$ order Bessel function.
Hence,
\begin{eqnarray}
  e^{\Phi^2\cos(\Omega t)}&=&J_0(-i\Phi^2)+2\sum_{n=1}^\infty i^nJ_n(-i\Phi^2)\cos(n\Omega t)\nonumber\\
                          &=&I_0(-\Phi^2)+2\sum_{n=1}^\infty i^n i^n I_n(-\Phi^2)\cos(n\Omega t)\nonumber\\
  &=&I_0(\Phi^2)+2\sum_{n=1}^\infty I_n(\Phi^2)\cos(n\Omega t)
\end{eqnarray}
where $I_n$ is the $n^\textrm{th}$ order modified Bessel function, and we have used the identities $J_n(ix)=i^n I_n(x)$ and $I_n(-x)=(-1)^n I_n(x)$.

\section{Ensemble simulation comparison}
\label{sec:ensemble-simulation}
We simulate the process using the Euler--Maruyama method with $\delta t=1\,\textrm{ns}$, and sampled at $10\,\textrm{MS}/\textrm{s}$ for $1\,\textrm{s}$ to generate $10\,\textrm{M}$ points.
(The crudeness of the numerical method makes necessary the short time-step.)
For an ensemble of trajectories ($\sim 40$), we estimate the spectrum with Bartlett's method as described in the main text using $M=2^{16}$, and find the maximum likelihood estimate of the parameters, $\bm{\alpha}_\textrm{MLE}$.

We use $\Omega=2\pi\times 100\,\textrm{kHz}$, $\Gamma=(10,20,50,100)\,\textrm{kHz}$, and use $\Phi=0.1,0.2,0.5,1.0$ for calculating the phase-modulated spectrum.
We use a temperature of $300\,\textrm{K}$ and a mass of $10^{-18}~\textrm{kg}$; these are typical experimental values, and only their ratio enters the simulation.
In keeping with the experimental system, we use a centre frequency $f_0=3\,\textrm{MHz}$.
Since the physical parameters are encoded in the {\it shape} of the spectrum, and are not reliant on any scaling, we use a unity amplitude sinusoid.

For this simulation we expect full agreement across the spectrum.
For the experimental system, regions between the peaks are slightly polluted with either second-order transverse motional peaks, or cross-coupling terms.
Therefore, for low $\Gamma$ where these additional unmodeled peaks become visible, it becomes necessary to window the spectrum.
This has little effect on the probability density, since these regions, at low $\Gamma$, do not depend strongly on the model parameters, and so have little influence (other than an inconsequential constant offset) on the Whittle log-likelihood.
We explored this windowing for low $\Gamma$, but found our experiment limited by intensity noise, as described in Section~\ref{sec:RIN}, and so have not used any selective windowing in the work presented.

\vfill

\end{document}